\documentclass[nofootinbib,prd]{revtex4}

\usepackage{amsmath}
\usepackage{amsfonts}
\usepackage{amssymb}
\usepackage{graphicx}%
\usepackage{mathrsfs}
\usepackage{yfonts}
\usepackage{tipa}
\usepackage{bm}
\usepackage{bbm}
\usepackage{hyperref}
\usepackage{latexsym}
\usepackage{xcolor}
\usepackage[iso-8859-7]{inputenc}
\begin{document}
\title{Charged rotating Casimir wormholes}
\author{Remo Garattini}
\email{remo.garattini@unibg.it}
\affiliation{Universit\`{a} degli Studi di Bergamo, Dipartimento di Ingegneria e Scienze
Applicate,Viale Marconi 5, 24044 Dalmine (Bergamo) Italy and }
\affiliation{  I.N.F.N. - sezione di Milano, Milan, Italy.}
\author{Athanasios G. Tzikas}
\email{athanasios.tzikas@unibg.it}
\affiliation{Universit\`{a} degli Studi di Bergamo, Dipartimento di Ingegneria e Scienze
Applicate,Viale Marconi 5, 24044 Dalmine (Bergamo) Italy and }

\begin{abstract}

We investigate the conditions under which a rotating traversable wormhole can be supported by a Casimir source in the presence of an external electric field. Extending previous studies of static Casimir wormholes and neutral rotating configurations, we construct an electrically charged rotating Casimir wormhole solution and determine the  thermal stress-energy tensor required to consistently satisfy the Einstein field equations. A particularly simple configuration arises when the rotation is constant and coincides with that measured by a zero-angular-momentum observer (ZAMO). In this case, the rotating wormhole preserves the same redshift and shape functions as the well-known static charged Casimir case, provided that the angular velocity and thermal components satisfy specific constraints imposed by the field equations. We also examine a configuration in which the angular velocity depends on the radial coordinate and decreases exponentially away from the throat. This damping mechanism removes the unrealistic persistence of frame dragging at large distances, while still allowing a consistent solution supported by Casimir, electromagnetic and thermal contributions.

\end{abstract}
\maketitle

\section{Introduction}
\label{sec:intro}

Traversable wormholes (TW) are theoretical solutions predicted by general relativity and have been a topic of scientific interest since as early as 1957 \cite{Whe57}. At that time, Wheeler hypothesized that the fabric of spacetime at extremely small scales$-$referred to as \textit{spacetime foam}$-$might be capable of generating Planck-sized TW through quantum fluctuations of spacetime itself. Almost 30 years later, Morris and Thorne \cite{MoT88} proposed the existence of static traversable wormholes of any size, requiring that these objects are threaded by exotic matter of negative energy density at the throat, which violates the known energy conditions. Following this proposal, numerous attempts have been made to explore the structure of such spacetime bridges by incorporating additional degrees of freedom, including minimally and non-minimally coupled scalar fields \cite{Kim98,BaV00,GaL07}, electromagnetic fields \cite{ScA96,KiL01,Can24}, and Casimir sources \cite{Gar19,Gar23,GaT24}. Other interesting approaches include wormhole solutions formulated in a de Sitter background \cite{DMS18}, as well as configurations inspired by Einsteinian cubic gravity \cite{LYM25} and teleparallel gravity \cite{HGSB22,REF1}. Additionally, models incorporating the Generalized Uncertainty Principle (GUP) within the framework of modified $f(Q)$ or $f(R)$ gravity \cite{HGSB23,REF4}, and those arising from extended theories of gravity \cite{FBCL21, FBCL21b}, offer compelling alternatives that support traversable geometries under modified gravitational dynamics.
Furthermore, non-local gravity provides a framework in which traversable wormholes can be supported without requiring exotic matter~\cite{REF2}, while wormhole geometries have also been obtained in Ricci-inverse gravity through the inclusion of novel geometric contributions~\cite{REF3}. More recently, configurations involving topological defects such as global monopoles have been shown to enhance the stability and physical viability of wormhole solutions~\cite{REF5}.

Among the family of traversable wormholes,  notable examples are those involving rotation \cite{Teo98,PB:2000ms,Kuh03,Kim:2004ph,MN,AYA,KS,KS1,AA,KK,JCJ,CMV,TCCM}. These configurations are stationary and axially symmetric solutions of the Einstein Field Equations (EFE) that reduce to the conventional Morris-Thorne line element in the non-rotating limit. Recently, there have been efforts to examine the  conditions required for the existence of  rotating Casimir wormholes \cite{GaT24b} and to better understand the properties of the Stress-Energy Tensor (SET) that describe these structures. Let us also note that, in such a framework, the existence of an additional SET, also known as the thermal  tensor \cite{Hay99}, is a necessary component for having a consistent set of field equations.  The introduction of the thermal  tensor is motivated by relativistic thermodynamic considerations, where microscopic random motions of the matter fields can generate effective anisotropic stresses. In this sense, the thermal contribution does not represent an independent physical source of energy, but rather serves as an effective description of backreaction effects within the wormhole geometry, as long as the thermal energy density component is zero. In particular, it allows us to maintain consistency of the Einstein field equations and ensures that the combined Casimir, electromagnetic, and thermal components satisfy the requirements for traversability without introducing additional macroscopic energy densities. In contrast to other approaches where supplementary stress-energy contributions are introduced in an ad hoc manner \cite{Kim98,BaV00}, the thermal tensor provides a more physically motivated effective description of such corrections.

The references discussed above outline the broader framework of Casimir-supported wormholes, which are particularly appealing as they rely on a physically motivated source of negative energy, namely the Casimir effect. However, to the best of our knowledge, a configuration that simultaneously incorporates electric charge and rotation within this framework has not been constructed so far, nor has a solution been presented that consistently reduces to the corresponding static charged case in the limit of vanishing angular momentum.

In this paper, we extend the analyses of \cite{Gar23} and \cite{GaT24b} by considering a rotating traversable wormhole that includes an additional stress-energy tensor, beyond the Casimir and thermal contributions, in order to account for the presence of an external electric field. The resulting solutions describe electrically charged rotating Casimir wormholes. More specifically, in Sec.~\ref{sec:2}, we outline the fundamental elements of the rotating metric, along with the matter/energy sources that describe our system. In Sec.~\ref{sec:3}, we solve the EFE by forcing the redshift and shape functions of the rotating manifold to match those of the static charged Casimir wormhole~\cite{Gar23}, ensuring this way consistency between the two configurations when  rotation stops. The field equations then determine the allowed angular velocity function $\Omega(r)$ as well as the explicit form of the thermal stress-energy tensor required to support these functions.  In Sec.~\ref{sec:ExpOmega}, we explore a potential wormhole solution in which the rotation undergoes an exponential decrease as one moves away from the throat. This represents a more realistic scenario, since the exponential damping suppresses the rotation at infinity, leaving a highly rotating spacetime only in the vicinity of the wormhole throat. Finally, in Sec.~\ref{sec:conclusion}, we draw our conclusions.   Throughout this work, we use natural units by setting $\hbar = c = 1$ ($G=\ell_{\rm P}^2=M_{\rm P}^{-2}$) and $\varepsilon_0 = 1/4\pi\,$. Furthermore, the fine-structure constant is given by $\alpha=e^2\approx 1/137\,$.

\section{The metric and the Stress-Energy Tensor}
\label{sec:2}

In this section, we discuss the metric and the SET characterizing a rotating TW. The rotating metric is described by the following stationary and axially symmetric line element \cite{Teo98}
\begin{equation} \label{metric1}
\mathrm{d}s^{2}=-e^{2\Phi\left(  r,{\theta}\right)  }\,\mathrm{d}t^{2}+\frac{\mathrm{d}r^{2}
}{1-b(r,{\theta})/r}+r^{2}K^{2}(r,{\theta})\left[  \mathrm{d}\theta^{2}+\sin^{2}
{\theta}\,\left(  \mathrm{d}\varphi-\Omega(r,{\theta})\mathrm{d}t\right)  ^{2}\right]
\,,
\end{equation}
where $\Phi (  r,{\theta} )$ and $b(r,{\theta})$ denote the  redshift and shape functions respectively. The functions $K(r,{\theta})$ and $\Omega(r,{\theta})$ are arbitrary functions of $r$ and $\theta\,$, associated with the proper distance and the angular velocity of the wormhole. One can rearrange the above line element into the following form
\begin{equation} \label{metric2}
\mathrm{d} s^{2}=g_{tt}\, \mathrm{d} t^{2}+g_{rr} \mathrm{d} r^{2}+2g_{t\varphi} \mathrm{d}t \mathrm{d}\varphi+g_{{\theta\theta}%
}\mathrm{d}\theta^{2}+g_{\varphi\varphi}\,\mathrm{d}\varphi^{2}\,,
\end{equation}
with
\begin{align} \label{metric_potentials}
g_{tt} &  =-\left(  e^{2\Phi (  r,{\theta})  }-r^{2} K^{2}(r,{\theta})\sin^{2}{\theta}\Omega^{2}(r,{\theta})\right)  \,\\
g_{rr} &  =\frac{1}{1-b(r,{\theta})/r} \,\\
g_{t\varphi} &  =-r^{2}K^{2}(r,{\theta})\sin^{2}{\theta}\Omega(r,{\theta}) \,\\
g_{{\theta\theta}} &  =r^{2}K^{2}(r,{\theta}) \,\\
g_{\varphi\varphi} &  =r^{2}K^{2}(r,{\theta})\sin^{2}{\theta} \,.
\end{align}
In the non-rotating limit ($\Omega(r,{\theta}) \rightarrow 0$), the above metric reduces to the standard  metric of a static TW, given by
\begin{equation} \label{metric3}
\mathrm{d}s^{2}=-e^{2\Phi\left(  r\right)  }\,\mathrm{d}t^{2}+\frac{\mathrm{d}r^{2}}{1-b(r)/r}%
+r^{2}\,(\mathrm{d}\theta^{2}+\sin^{2}{\theta}\,\mathrm{d}\varphi^{2}) \,.
\end{equation}
The validity of this limit is crucial for our investigation, as the form of a static Casimir wormhole inspired by an electric source is a well-established solution \cite{Gar23}. In other words, the exact forms of the redshift and shape functions are known in the static limit and are given by
\begin{equation}\label{shape_red}
b(r) = r_0 \left( 1- \frac{1}{\omega} \right)  +  \frac{r_0^2}{\omega r} \qquad \mathrm{and} \qquad \Phi(r) = \frac{\omega-1}{2} \ln \left( \frac{\omega r}{\omega r + r_0} \right) \,,
\end{equation}
where the constant $\omega$ reads
\begin{equation}\label{w}
\omega =  \frac{r_0^2}{r_1^2-r_2^2} \,.
\end{equation}
These forms must also hold in the rotating frame to ensure that the solution reduces consistently to the static limit. The value $r=r_0$ denotes the throat of the wormhole, while the lengths $r_1$ and $r_2$ are defined as
\begin{equation} \label{r1_r2}
 r_1^2 = \frac{\pi^3 \ell^2_{\mathrm{P}}}{90} \qquad \mathrm{and} \qquad r_2^2 =  Q^2 \ell^2_{\mathrm{P}} \,,
\end{equation}
with $Q$ being the electric charge of the source. The conventional charge quantization $Q=N e = \frac{N}{\sqrt{137}}$ forbids the exact equality of the above two radii ($r_1\neq r_2$). Specifically, if we set $r_1=r_2\,$, we find $N = \sqrt{\frac{137 \pi^3}{90}}\approx 6.87\,$, which is invalid since $N $ must be a positive integer. Consequently, for $N \leq 6$ we get $r_1>r_2$ and $\omega >0\,$, whereas for $N \geq 7\,$, we get $r_2>r_1$ and $\omega <0\,$. 

Furthermore, the functions \eqref{shape_red}  must align with those of the rotating metric to ensure consistency with the static limit. As a result, the angular dependence is absent from these functions, i.e., $\Phi(  r,{\theta}) \equiv \Phi(  r)$ and $b(  r,{\theta}) \equiv b( r )\,$. Following this reasoning, the dimensionless function $K(r,\theta)$ can be set equal to unity ($K(r,\theta)=1$) without  loss of generality. Accordingly, we aim to investigate whether this electrically charged solution can also exist in a rotating frame by determining the form of the remaining  function  $\Omega(r,{\theta})$, as well as the components of the SET required to complete the solution. 

Next, we focus on the anisotropic form of the SET describing our system, which can be decomposed into three parts; a Casimir part $T^{\mu}_{\nu}|_{\rm C}$, an electric part $T^{\mu}_{\nu}|_{\rm E}$ and a thermal part $T^{\mu}_{\nu}|_{\rm Th}$  (thermal stress tensor). These are expressed as follows:
\begin{align} \label{SET_cas}
 T^{\mu}_{\nu}|_{\rm C}  &   =\mathrm{diag}\left[ - \rho_{\mathrm{C}}(r) , \, p_{\mathrm{r,C}}(r)  , \, p_{\mathrm{t,C}}(r)  , \, p_{\mathrm{t,C}}(r)  \right]= - \frac{r_{1}^{2}}{\kappa r^{4}}\mathrm{diag}\left[  -1,3,-1,-1\right] \\
 T^{\mu}_{\nu}|_{\rm E}  & =\mathrm{diag}\left[ - \rho_{\mathrm{E}}(r)  , \, p_{\mathrm{r,E}}(r)  , \, p_{\mathrm{t,E}}(r)  , \, p_{\mathrm{t,E}}(r)  \right]  = \frac{r_2^{2}}{ \kappa  r^{4} } \mathrm{diag} \left[ -1, -1, 1, 1 \right] \label{SET_em} \\
 T^{\mu}_{\nu}|_{\rm Th}  &  = \mathrm{diag}\left[ - \tau_{\rho}(r)  , \,  \tau_{\rm r}(r)  , \,  \tau_{\rm t}(r)  , \, \tau_{\rm t}(r)  \right] \,. \label{SET_th}
\end{align}
The components $\rho_{i}$, $p_{\mathrm{r},i}$ and $p_{\mathrm{t},i}$ (with $i=\mathrm{C,E}$) represent the energy density, radial pressure, and tangential pressure, respectively, for each part. The expression  in \eqref{SET_cas} corresponds to  the SET of a Casimir apparatus with radially variable conducting plates. 
In addition, the thermal stress tensor has been decomposed into an
energy component $\tau_{\rho}(r)$, a radial component $\tau_{\rm r}(r)$ and a tangential component $\tau_{\rm t}(r)\,$. We may express the total SET in the following form
\begin{equation}
T_{\mu\nu}=\left(  \rho(r)+\tau_{\rho}(r)\right)  u_{\mu}u_{\nu}+\left(  p_{\rm r}(r)+\tau_{\rm r}(r)\right)  n_{\mu}n_{\nu}+\left(  p_{\rm t}(r)+\tau_{\rm t}(r)\right)  \sigma_{\mu\nu}\,,
\label{SET}
\end{equation}
where the unit vectors $u_{\mu}$ and $n_{\mu}$ are  timelike and spacelike, respectively, and the operator
\begin{equation}
\sigma_{\mu\nu}=g_{\mu\nu}+u_{\mu}u_{\nu}-n_{\mu}n_{\nu}%
\end{equation}
is a projection operator onto a two-surface orthogonal to $u_{\mu}$ and $n_{\mu}\,$. Here, we  define
\begin{align}
 \rho(r)&=\rho_{\rm C}(r)+\rho_{\rm E}(r) = - \frac{r_1^2-r_2^2}{\kappa r^4} = - \frac{r_0^2}{\omega \kappa r^4}    \label{e1} \\
 p_{\rm r}(r)&=p_{\rm r, C} (r)+p_{\rm r, E}(r) = - \frac{3r_1^2+r_2^2}{\kappa r^4} \label{e2}\\
 p_{\rm t}(r)&=p_{\rm t, C}(r)+p_{\rm t, E}(r) = \frac{r_1^2+r_2^2}{\kappa r^4}  \,. \label{e3}
\end{align}
To incorporate rotations, we utilize the Killing vectors $\xi_{t}^{\alpha}=\delta_{t}^{\alpha}$ and $\xi_{\varphi}^{\alpha}=\delta_{\varphi}^{\alpha}\,$. 
Based on the reasoning in \cite{GaT24},  $u^{\mu}$ can be expressed as
\begin{equation}
u^{\mu}=\frac{e^{-\Phi (  r )  }}{\sqrt{1-v^{2}}  }\left(  1,0,0,\Omega_{0}\right)  \,,
\end{equation}
where 
\begin{equation}
v=r \sin{\theta}\left(  \Omega(r,{\theta})-\Omega_{0}\right) \,e^{-\Phi (  r )  }%
\end{equation}
is the proper velocity of the matter with respect to a ZAMO. The ZAMO frame plays a central role in the physical interpretation of the rotating wormhole geometry. Observers in this frame have vanishing angular momentum with respect to infinity and are locally non-rotating, effectively absorbing the frame-dragging effects induced by the rotating spacetime. As a result, the angular velocity measured in this frame corresponds to the locally inertial rotation of spacetime itself. In the present model, adopting the ZAMO frame significantly simplifies the Einstein field equations and provides a natural setting in which the rotating solution consistently reduces to the static charged Casimir wormhole when the angular velocity vanishes. The angular velocity $\Omega_0$ measured by a distant observer is given by
$
\Omega_{0}= u^{\varphi}/u^{t} 
$. 
Using the above information, the components of the total SET in the rotational frame can be expressed as
\begin{align}
T_{tt} & = \left[  \rho(r)+\tau_{\rho}(r) +p_{\rm t}(r)+\tau_{\rm t}(r)\right]  u_{t}u_{t}+\left[  p_{\rm t}(r)+\tau_{\rm t}(r)\right]
g_{tt} \label{Ttt}\\
T_{rr} &  =\left[  p_{\rm r}(r) +\tau_{\rm r}(r) \right]  g_{rr} \label{Trr}\\
T_{{\theta\theta}} & = \left[  p_{\rm t}(r) + \tau_{\rm t}(r)\right]  g_{{\theta\theta}} \label{Tthth}\\
T_{\varphi\varphi} & =\left[  \rho(r)+\tau_{\rho}(r)+p_{\rm t}(r)+\tau_{\rm t}(r)\right]  u_{\varphi}u_{\varphi}+\left[  p_{\rm t}(r)+\tau_{\rm t}(r)\right] g_{\varphi\varphi} \label{Tpp}\\
T_{t\varphi} & = T_{\varphi t} =  \left[  \rho(r) +\tau_{\rho}(r)+p_{\rm t}(r)+\tau_{\rm t}(r)\right]  u_{t}u_{\varphi}+\left[  p_{\rm t}(r)+\tau_{\rm t}(r)\right]  g_{t\varphi} \,\label{Ttp}%
\end{align}
with
\begin{equation}
u_{t}   = \frac{e^{-\Phi (  r )  }}{\sqrt{1-v^{2}}  }\left(  g_{tt}+\Omega_{0}g_{t\varphi}\right)  
\quad \mathrm{and} \quad u_{\varphi}   = \frac{e^{-\Phi (  r )  }}{\sqrt{1-v^{2}}  } \left(  g_{t\varphi}+\Omega_{0}g_{\varphi
\varphi}\right)  .
\end{equation}
The thermal parameters $\tau_{\rho}(r)\,$,   $\tau_{\rm r}(r)$ and $\tau_{\rm t}(r)\,$, along with the rotation parameter $\Omega(r,{\theta})\,$, have to be determined.

\section{Rotation and thermal parameters}
\label{sec:3}

Our goal is to determine the rotation and thermal parameters. The components of the SET are given by \eqref{e1}-\eqref{e3}, along with the components of the thermal tensor. Before examining the non-vanishing components of the EFE, we first check the constrain arising from the vanishing of the $\theta r-$component:
\begin{equation}
G_{r\theta}=\frac{r^{2}}{2}\sin^{2}\left(  \theta\right)  e^{-2\Phi\left(
r\right)  }\frac{\partial\Omega\left(  r,\theta\right)  }{\partial\theta
}\frac{\partial\Omega\left(  r,\theta\right)  }{\partial r}=0 \,.
\end{equation}
This field equation is satisfied if
\begin{equation}
\Omega\left(  r,\theta\right)  \rightarrow\Omega\left(  r\right) \qquad \mathrm{or} \qquad \Omega\left(  r,\theta\right)  \rightarrow\Omega\left(  \theta\right) \,.
\end{equation}
In this paper, we  choose the form  $\Omega(r,\theta) \equiv \Omega(r)\,$.
We proceed now with the $rr-$component of the EFE, which reads
\begin{equation}
- \frac{r_0^2}{r^4} + \frac{r (r-r_0) \sin^2\theta \ \Omega '(r)^2 }{4}  \left( \frac{r_0+\omega r}{\omega r} \right)^{\omega}  = - \frac{3r_1^2+r_2^2}{r^4} + \kappa \tau_{\rm r}(r) \,.
\end{equation}
Examining the throat ($r=r_0$), we get
\begin{equation}
\tau_{\rm r}(r_0) = \frac{3r_1^2+r_2^2-r_0^2}{\kappa r_0^4}
\end{equation}
and a simplified solution  occurs for $\tau_{\rm r}(r_0)=0$ if
\begin{equation} \label{r0r1r2}
3r_1^2+r_2^2=r_0^2 \,.
\end{equation}
Using the relations  \eqref{w} and \eqref{r1_r2}, we get
\begin{equation} \label{r1r2}
r_2^2 = \frac{\omega-3}{\omega+1} r_1^2 \quad \Longrightarrow \quad N^2 = \frac{137\pi^3}{90} \left( \frac{\omega-3}{\omega+1} \right) \,
\end{equation}
and, since $N^2>0\,$, the range of $\omega$ is restricted as $\omega>3$ or $\omega<-1\,$. We may also write the two characteristic lengths with respect to the throat as
\begin{equation} \label{r012}
r_1^2 = \frac{\omega+1}{4\omega} r_0^2 \qquad \mathrm{and} \qquad r_2^2= \frac{\omega-3}{4\omega} r_0^2 \,.
\end{equation}
Note that for the special value $\omega$=3, one recovers the original relationship between the throat radius and the Planck length.
Taking the above condition into consideration, we may write the general form for $\tau_{\rm r}(r)$ as
\begin{equation}
\tau_{\rm r}(r) = \frac{r (r-r_0) \sin^2\theta \ \Omega '(r)^2 }{4\kappa}  \left( \frac{r_0+\omega r}{\omega r} \right)^{\omega} .
\end{equation}
It is obvious that for a constant rotation ($\Omega(r)=\Omega$) or at the throat,  the radial thermal component vanishes ($\tau_{\rm r}(r)=0$).

We move on to the $\theta\theta-$component of the EFE, which is given by
\begin{equation}
\frac{r_0^2 (r_0+4\omega r_0+4\omega^2 r-\omega^2r_0)}{4\omega r^4 (r_0+\omega r)}- \frac{r (r-r_0) \sin^2\theta \ \Omega '(r)^2 }{4}  \left( \frac{r_0+\omega r}{\omega r} \right)^{\omega} = \frac{r_1^2+r_2^2}{r^4} + \kappa \tau_{\rm t}(r) \,.
\end{equation}
For $\Omega(r)=\Omega$ and upon using \eqref{r0r1r2}, \eqref{r1r2} and \eqref{r012}, we get the form of the tangential thermal component
\begin{equation} \label{Tt1}
\tau_{\rm t}(r) = \frac{r_0^2 (r_0+4\omega r_0+4\omega^2 r-\omega^2r_0)}{4\kappa\omega r^4 (r_0+\omega r)} - \frac{r_1^2+r_2^2}{\kappa r^4} = \frac{2r_1^2}{\kappa r^4} \left( \frac{\omega r+ (\omega - 3)r_1 \sqrt{\frac{\omega}{1+\omega}}}{\omega r+ 2r_1 \sqrt{\frac{\omega}{1+\omega}}} \right) 
\end{equation}
and in the vicinity of the throat it takes the value
\begin{equation} \label{Ttr0}
\tau_{\rm t}(r_0) = \frac{3r_0^2 - 3r_1^2-5r_2^2}{4\kappa r_0^4} = \frac{3r_1^2-r_2^2}{2 \kappa (3r_1^2+r_2^2)^2} = \frac{\omega+3}{4\kappa \omega r_0^2} = \frac{(1+\omega) (3+\omega)}{16\kappa \omega^2 r_1^2} \,.
\end{equation}
The expression \eqref{Tt1} simplifies when $\omega$ takes the permitted value $\omega = 5\,$. In this case, $r_1^2 = 3r_2^2\,$, and the thermal component reduces to
\begin{equation} \label{Tt1b}
\tau_{\rm t}(r) = \frac{2r_1^2}{\kappa r^4}
\end{equation}
with
\begin{equation}
\tau_{\rm t}(r_0) = \frac{3}{25\kappa r_1^2} \,.
\end{equation}
Under this condition, the number of elementary charges is $N \approx 4$ (to be precise $\omega=5.0514$ so that $N=4$) and the throat is of the order
\begin{equation}
r_0 = 0.73 \ell_{\rm P} \,.
\end{equation}
Next, we examine the EFE $G^{\varphi}_{\varphi}= \kappa T^{\varphi}_{\varphi}\,$. The right-hand side (RHS) is
\begin{equation} \label{Tff}
\kappa T^{\varphi}_{\varphi}= \frac{r_1^2+r_2^2}{r^4} + \kappa \tau_{\rm t}(r) + \frac{\left( \frac{2r_2^2}{ r^4}+ \kappa \tau_{\rm t}(r)+\kappa \tau_{\rho}(r)\right)  (\Omega_0-\Omega)  \Omega_0 \sin^2\theta}{  (\Omega-\Omega_0)^2  \sin^2\theta - \frac{1}{r^2} \left( \frac{\omega r}{\omega r + r_0}\right)^{\omega-1}   }\,,
\end{equation}
while the left-hand side (LHS) is rather lengthy but  can be simplified once $\Omega=\mathrm{const.}$:
\begin{equation} \label{Gff}
G^{\varphi}_{\varphi} = \frac{r_0^2 (r_0+4\omega r_0+4\omega^2 r-\omega^2r_0)}{4 \omega r^4 (r_0+\omega r)} \,.
\end{equation}
The specific field equation is satisfied in two cases. First, when working within the ZAMO frame ($\Omega=\Omega_0$). In this case, by equating \eqref{Tff} with \eqref{Gff}, yields an expression for $\tau_{\rm t}(r)$ that matches the expression given by  \eqref{Tt1} or \eqref{Tt1b}. Second, we may have $\Omega\neq \Omega_0$ but the condition
\begin{equation} \label{Te1a}
\tau_{\rho}(r) = - \tau_{\rm t}(r) - \frac{2r_2^2}{\kappa r^4} 
\end{equation}
must hold. This ensures that the third term of \eqref{Tff} vanishes, preserving the form of $\tau_{\rm t}(r)$ as previously derived.

The next EFE corresponds to the $tt-$component, expressed as $G^t_t=\kappa T^t_t\,$. The RHS  of the field equation is 
\begin{equation} \label{efe3}
\kappa T^t_t = \frac{r_1^2+r_2^2}{r^4} + \kappa \tau_{\rm t}(r) - \frac{\left( \frac{2r_2^2}{ r^4}+ \kappa \tau_{\rm t}(r)+\kappa \tau_{\rho}(r)\right)\left( 1- r^2 \sin^2\theta   (\Omega-\Omega_0) \Omega  \left( \frac{\omega r}{\omega r + r_0}\right)^{1-\omega} \right) }{\left( 1- r^2 \sin^2\theta   (\Omega-\Omega_0)^2   \left( \frac{\omega r}{\omega r + r_0}\right)^{1-\omega} \right)}\,,
\end{equation}
while the LHS simplifies to
\begin{equation}
G^t_t = \frac{r_0^2}{\omega r^4}
\end{equation}
by assuming a constant rotational parameter  $\Omega\,$. Substituting the expression \eqref{Te1a} for $\tau_{\rho}(r)$ into \eqref{efe3}, we obtain the following form for the tangential component
\begin{equation} \label{Tta2}
\tau_{\rm t}(r) = - \frac{2r_2^2}{\kappa r^4}
\end{equation}
which contradicts \eqref{Tt1b}. Therefore, the solutions \eqref{Te1a} and \eqref{Tta2} are  discarded. The second choice is the ZAMO reference frame. In this frame, the  field equation \eqref{efe3} becomes
\begin{equation}
\frac{r_0^2}{\omega r^4} = \frac{r_1^2+r_2^2}{r^4} + \kappa \tau_{\rm t}(r)- \left( \frac{2r_2^2}{ r^4}+ \kappa \tau_{\rm t}(r)+\kappa \tau_{\rho}(r)\right)
\end{equation}
and upon utilizing \eqref{w}, we get
\begin{equation}
\tau_{\rho}(r) =0 \,.
\end{equation}
As mentioned in the introduction, the vanishing of $\tau_{\rho}$ leads to a simplified model with fewer external parameters and ensures that the solution is not influenced by additional energy sources associated with the thermal tensor. The remaining thermal pressure components can instead be interpreted as a backreaction of the wormhole geometry to the underlying matter configuration.
There are two remaining EFE components to examine; the $t\varphi-$component and the $\varphi t-$component. Although the covariant and contravariant forms of the Einstein tensor and the SET remain unchanged when swapping the indices, i.e.,
\begin{equation}
G_{t\varphi} = G_{\varphi t}\,,  \qquad T_{t\varphi} = T_{\varphi t} \,,
\end{equation}
and
\begin{equation}
G^{t\varphi} = G^{\varphi t} \,, \qquad T^{t\varphi} = T^{\varphi t} \,,
\end{equation}
 the mixed terms differ ($G^t_{\varphi} \neq G^{\varphi}_t$) due to the distinct metric components $g_{tt}$ and $g_{\varphi\varphi}$ used when raising or lowering indices. The EFE $G^t_{\varphi}= \kappa T^{t}_{\varphi}$ is satisfied in the ZAMO frame by applying the previously derived expressions for the thermal parameters, since both the Einstein tensor and the corresponding SET vanish identically. As for the final EFE $G_t^{\varphi}= \kappa T_{t}^{\varphi}\,$,  in the  ZAMO frame, the RHS becomes
\begin{equation} \label{efeft}
\kappa T_{t}^{\varphi} = - \left(  \frac{2r_2^2}{r^4} + \kappa \tau_{\rm t}(r) \right)  \Omega_0\,,
\end{equation} 
where $\tau_{\rm t}(r)$ is given by the general solution \eqref{Tt1}. The LHS of the EFE  reads
\begin{equation}
G^{\phi}_t = \frac{r_0^2 (\omega - 1)(r_0(\omega-3)-4 \omega r ) \Omega_0}{4\omega r^4(\omega r + r_0)}
\end{equation}
and is identical to \eqref{efeft} after applying \eqref{r012} and \eqref{Tt1}. Furthermore, it is evident that the null energy condition (NEC) is violated, since the absence of the radial thermal component leads to the relation
\begin{equation} \label{nec}
    \rho(r)+p_{\rm r}(r) = - \frac{r_0^2 (\omega+1)}{\omega \kappa r^4} <0 \,.
\end{equation}
The violation of the NEC is a key requirement for the traversability of the wormhole.

\section{Exponentially damped rotation}
\label{sec:ExpOmega}

In the previous section, we demonstrated that a constant rotation $\Omega=\Omega_0\,$, satisfies the Einstein field equations (EFE), resulting in a viable wormhole solution within the ZAMO frame. However, this solution possesses the unpleasant feature of maintaining rotation  at large distances, implying a non-vanishing frame-dragging effect even at infinity. To address this, we introduce an exponential damping of the rotation, expressed as 
\begin{equation} \label{DO}
\Omega\left(  r\right)  =\Omega \ e^{ -\mu\left(  r-r_{0}\right)} \,,
\end{equation}
where $\mu$ is the damping factor with dimensions of inverse length. Of course, in the limit $\mu \rightarrow \infty$, we recover the static case, since $\Omega \rightarrow 0$. The exponentially damped rotation profile is introduced to ensure a physically acceptable asymptotic behavior of the spacetime. In particular, it prevents the persistence of frame-dragging effects at large distances and guarantees that the geometry approaches a non-rotating configuration far from the throat. From a physical viewpoint, this corresponds to a situation in which the angular momentum of the system is effectively localized near the wormhole throat, rather than being distributed throughout the space. From a theoretical perspective, the damping introduces a characteristic length scale $\mu^{-1}$, which smoothly interpolates between a highly rotating near-throat region and an asymptotically static regime, thereby providing a controlled extension of the constant-rotation solution. It is our goal to examine whether a consistent set of parameters exists for the validity of the field equations. First of all, let us check the definition of the ergoregion where $g_{tt}=0$. Since we wish $g_{tt}<0$ for a positive-definite metric, we get the following inequality
\begin{equation}
\Omega(r) < \frac{1}{r \sin\theta} \left( \frac{\omega r}{\omega r +r_0} \right)^{\frac{\omega-1}{2}} .
\end{equation}
Numerical analysis shows that the RHS of the inequality is a monotonically decreasing function of the coordinate $r$ when $r>r_0\,$ and once $\theta$ is fixed. Therefore, the rotational parameter should decrease in value more rapidly as one moves away from the throat than what is suggested by the RHS of the above inequality. This motivates the definition given in \eqref{DO}. 
Starting again with the $rr-$component of the EFE, we get
\begin{equation}
- \frac{r_0^2}{r^4} + \frac{\mu^2 r (r-r_0) \sin^2\theta \ \Omega^2 }{4}  \left( \frac{r_0+\omega r}{\omega r} \right)^{\omega} e^{-2\mu(r-r_0)}  = - \frac{3r_1^2+r_2^2}{r^4} + \kappa \tau_{\rm r}(r) \,.
\end{equation}
In the vicinity of the throat, the relations \eqref{r0r1r2}-\eqref{r012} hold, leading to the same limits for $\omega$ and ensuring that the radial thermal component vanishes at the throat ($\tau_{\rm r}(r_0)=0$). Solving for $\tau_{\rm r}(r)\,$, we obtain a non-vanishing component
\begin{equation} \label{Tr2}
 \tau_{\rm r}(r)= \frac{\mu^2 r (r-r_0) \sin^2\theta \ \Omega^2 }{4\kappa}  \left( \frac{r_0+\omega r}{\omega r} \right)^{\omega} e^{-2\mu(r-r_0)}  \,,
\end{equation}
which contrasts with the constant rotation case, as it is non-zero everywhere except at the throat. Of course, the above expression vanishes for constant rotation ($\mu=0$) and at very large distances ($r \gg r_0$). Let us note that there is also a $\theta-$dependence in the expression, but it  can be ignored once we fix the equatorial plane ($\theta=\pi/2$). Otherwise, a corrective numerical coefficient arises for different $\theta-$planes. 

Switching to the $\theta\theta-$component of the EFE, we get the following field equation
\begin{equation}
\frac{r_0^2 (r_0+4\omega r_0+4\omega^2 r-\omega^2r_0)}{4\omega r^4 (r_0+\omega r)} - \tau_{\rm r}(r) = \frac{r_1^2+r_2^2}{r^4} + \kappa \tau_{\rm t}(r) \,,
\end{equation}
from which we can extract the form of the tangential thermal parameter 
\begin{equation} \label{Tt2}
\tau_{\rm t}(r)= \frac{r_0^2 (1+\omega) \left( 2\omega r+ r_0 (3-\omega) \right) }{4\kappa \omega r^4 (\omega r+r_0)}-  \tau_{\rm r}(r)   \,,
\end{equation}
after using \eqref{r012}.
At the throat, this component coincides with the expression \eqref{Ttr0}.
Next, the tensor $G^{t}_{t}$ is given by
\begin{equation} \label{Gtt2}
G^{t}_{t}=\frac{r_0^2}{\omega r^4} + A(r) B(r) \sin^2\theta \,,
\end{equation}
admitting the following form at the throat
\begin{equation}
G^{t}_{t} |_{r=r_0} = \frac{1}{\omega r_0^2} - \frac{1}{4}r_0 \Omega^2 \mu \left( \frac{1+\omega}{\omega} \right)^{\omega} \sin^2\theta \,.
\end{equation}
The expressions for the functions $A(r)$ and $B(r)$ are
\begin{align}
A(r) &= r_0 (7r_0-8r)-\omega (8r^2-8r_0 r +r_0^2)+3r \mu (r-r_0)(\omega r+r_0) \\
B(r) & =  \frac{ \Omega^2 \mu  \ e^{-2\mu(r-r_0)}}{4(\omega r+r_0)}  \left( \frac{\omega r+r_0}{\omega r} \right)^{\omega} .
\end{align}
The corresponding component of the SET reads
\begin{equation} \label{Ttt2}
\kappa T^t_t = \frac{r_1^2+r_2^2}{r^4} + \kappa \tau_{\rm t}(r) - \left( \frac{2r_2^2}{ r^4}+ \kappa \tau_{\rm t}(r)+\kappa \tau_{\rho}(r)\right) C(r) \,,
\end{equation}
where
\begin{equation} 
C(r) =   \frac{ 1- r^2 \sin^2\theta   (\Omega \ e^{ -\mu\left(  r-r_{0}\right)}-\Omega_0) \Omega \ e^{ -\mu\left(  r-r_{0}\right)}  \left( \frac{\omega r}{\omega r + r_0}\right)^{1-\omega}  }{1- r^2 \sin^2\theta   (\Omega \ e^{ -\mu\left(  r-r_{0}\right)}-\Omega_0)^2   \left( \frac{\omega r}{\omega r + r_0}\right)^{1-\omega} } \,.
\end{equation}
By equating \eqref{Gtt2} with \eqref{Ttt2}, we recover the form of the last thermal component
\begin{equation}
\tau_{\rho} (r) = \left( \frac{(\omega-3)r_0^2}{2\kappa \omega r^4} + \tau_{\rm t}(r) \right) \left(\frac{1}{C(r)}-1 \right) - \frac{A(r) B(r)}{\kappa C(r)} \sin^2\theta \,. 
\end{equation}
For a constant rotation ($\mu=0$), the above component vanishes ($\tau_{\rho} (r)=0$). In the vicinity of the throat, it takes the form of
\begin{equation}
\tau_{\rho} (r_0) = \frac{\sin^2\theta \left( r_0^3 \mu \left(\frac{\omega+1}{\omega}\right)^{\omega}\omega \Omega^2 (\Omega-\Omega_0)^2 \sin^2\theta)+3(1-\omega)(\Omega-\Omega_0)\Omega_0-\mu r_0(1+\omega)\Omega^2 \right)}{4\kappa \omega r_0^2 \Omega (\Omega-\Omega_0) \sin^2\theta -4\kappa \left(\frac{\omega}{1+\omega}\right)^{\omega} (1+\omega)}
\end{equation}
and, working in the ZAMO frame ($\Omega=\Omega_0$), it simply becomes
\begin{equation}
\tau_{\rho} (r_0) = \frac{1}{4\kappa}r_0 \Omega_0^2 \mu \left( \frac{1+\omega}{\omega} \right)^{\omega} \sin^2\theta \,.
\end{equation}
This contrasts with the analysis in \cite{GaT24b}, as in the neutral rotating case there is no need for an additional thermal energy density, whereas in the charged case such a contribution is required. Regarding the remaining field equations, which involve the $\varphi\varphi-$component and the mixed terms, they are satisfied if we substitute the previously derived expressions of the thermal tensor, provided the exponential damping factor $\mu$ is sufficiently small. Such a small value of $\mu$ implies an almost vanishing $\tau_{\rho}$, leaving the interpretation of the thermal tensor essentially unchanged, namely as a backreaction of the system arising from the remaining thermal pressure components. It is worth noting that the NEC is violated under such a small factor $\mu$ since it leads to the relation \eqref{nec}.

\section{Conclusions}
\label{sec:conclusion}

We have investigated the conditions necessary for forming a traversable wormhole in a rotating frame, considering a Casimir apparatus coupled with an external electric field. Our approach preserves the forms of the redshift and shape functions as in the static case, ensuring compatibility with the well-established geometry of a  charged Casimir wormhole in the non-rotating frame. This framework allows us to determine whether the thermal components and the rotation parameter can be appropriately constrained to recover the desired rotating solution.

By analyzing the SET of the system, we find that a rotating wormhole solution is feasible, provided the rotation is constant and the only non-vanishing thermal contribution is the tangential pressure component, as given by \eqref{Tt1}. However, constant rotation leads to frame dragging persisting even at spatial infinity. To resolve this, we introduce an exponential damping in the rotation parameter that decreases with radial distance from the throat. In this case, a consistent solution can be obtained if all components of the thermal SET are non-zero and the damping parameter is sufficiently small to satisfy the field equations. In contrast to the neutral case, this analysis requires the inclusion of a thermal energy density when exponential damping is applied, whereas it is absent in the uncharged rotating case. Last but not least, we must stress that the form we used for the shape function is valid in the static case only if an equation  of the form $p_{\rm r}(r)=\omega \rho(r)$ is imposed. In our rotating scenario, however, no such an equation  has been imposed. Instead, we have adopted this profile solely because it must remain valid when rotation ceases. In this sense, we have an asymptotic shape function (asymptotic with respect to rotation) that matches the rotating configuration. The rotational properties are fully captured by the rotation parameter $\Omega(r)$.
These results summarize a consistent construction of charged rotating Casimir wormholes and clarify the role of rotation and thermal effects in maintaining the viability of the solution. The model provides a controlled framework in which static configurations are recovered in the appropriate limit, while rotation introduces non-trivial modifications to the stress-energy content.

Possible extensions of this work include a detailed stability analysis of the rotating configurations, further generalizations of the rotation profile, and the study of additional matter sources that could support more realistic astrophysical scenarios. In particular, it would be interesting to extend the present framework to higher-dimensional spacetimes, where Casimir-type effects may acquire additional geometric contributions, or to modified gravity theories such as Einstein-Gauss-Bonnet gravity, where higher-curvature corrections could play a significant role in the wormhole structure. Another natural direction is the inclusion of non-linear electrodynamics, which is known to regularize field strengths and may lead to improved energy-condition behaviour in charged wormhole geometries. These extensions could provide further insight into the construction of physically viable traversable wormhole solutions.

\end{document}